# Harnessing label semantics to extract higher performance under noisy label for Company to Industry matching

Apoorva Jaiswal[†]
Machine Learning Engineer
JPMorgan Chase Commercial Banking
Bangalore, Karnataka, India
apoorva.jaiswal@jpmchase.com

Abhishek Mitra[†]
Machine Learning Engineer
JPMorgan Chase Commercial Banking
Bangalore, Karnataka, India
abhishek.x.mitra@chase.com

## ABSTRACT

Assigning appropriate industry tag(s) to a company is a critical task in a financial institution as it impacts various financial machineries. Yet, it remains a complex task. Typically, such industry tags are to be assigned by Subject Matter Experts (SME) after evaluating company business lines against the industry definitions. It becomes even more challenging as companies continue to add new businesses and newer industry definitions are formed. Given the periodicity of the task it is reasonable to assume that an Artificial Intelligent (AI) agent could be developed to carry it out in an efficient manner. While this is an exciting prospect, the challenges appear from the need of historical patterns of such tag assignments (or Labeling). Labeling is often considered the most expensive task in Machine Learning (ML) due its dependency on SMEs and manual efforts. Therefore, often, in enterprise set up, an ML project encounters noisy and dependent labels. Such labels create technical hindrances for ML Models to produce robust tag assignments. We propose an ML pipeline which uses semantic similarity matching as an alternative to multi label text classification, while making use of a Label Similarity Matrix and a minimum labeling strategy. We demonstrate this pipeline achieves significant improvements over the noise and exhibit robust predictive capabilities.

## 1 Introduction

Assigning industry to a company is critical, yet error prone process. Industry definitions evolve over time as companies continue to grow their business portfolios, therefore needing a continuous assignment process. This generates an opportunity to develop a text classification ML algorithm which intakes a textual description of a company's business and outputs one or more appropriate industry classes (or tags). Yet, it meets with (1) "cold start" as well as (2) "operational" challenges. The primary "cold start" issue arises from the need of historical pattern of industry assignments by SMEs.

[†]Equal Contributors

Such historical patterns are often (1.a) Partial and/or (1.b) Incorrect. Illustrations of such assignments are given in table 1.

| Company Business Description (CBD) | Industry Tags by Labelers | Gold Standard Tags | Accuracy |
|---|---|---|---|
| Desc1 | I1, I2 | I1, I2, I3 | Partial |
| Desc2 | I1, I2, I3 | I1, I2 | Partial |
| Desc3 | I1, I2 | I3 | Incorrect |

Table 1: Original dataset format

The (2) "operational" challenge arises from the need of human SMEs/Labelers to do this assignment periodically. Therefore, our work focuses on developing an ML pipeline to assign appropriate industry tags (TAG) to company business descriptions (CBD) learning from a historical pattern of partial or incorrect tags assignments and with minimal labeling corrections i.e., Multi Label Classification (MLC) with noisy labels. The **novelty** of our work lies in A) **Using semantic similarity matching as an alternative to multi label text classification with noisy labels** B) **Integrate label dependencies into semantic similarity model through a rated Label Similarity Matrix (LSM)** C) **Reducing human effort through minimum labeling strategy and** D) **Building a single model that can perform both A) and B)**. We compare our results with SME curated gold standard assignments to demonstrate the improvement over the noise. One of the highlight achievements is getting robust performance where labels are extremely noisy.

## 2 Related Work

The technical problem can be deconstructed into two main subsections; (2.1) Multi Label Text Classification [MLC] [1][2] and (2.2) Text Classification under Noisy Labels.

*2.1:* Broadly there are two approaches to MLC, e.g., **Problem Transformation Methods** and **Algorithm Adaptation Methods** [1]. Problem Transformation Methods either transform the MLC into a set of Multi Class Classification problems (e.g., Binary Relevance) or transform the original label combinations into unique synthetic labels (e.g., Label Powerset). The challenges with such methods are well documented [1]. Algorithms similar to Binary Relevance suffer from high model complexities and high inference times [1]. Label Powerset and similar algorithms suffer from generalization errors [1]. While algorithm adaptation methods improve upon some of these shortcomings still it encounters label confusion errors, high time complexities and scaling challenge over large number of labels [1]. The aforesaid MLC algorithms also suffer from a particular issue of **Label Dependencies** [1]. Labels can be semantically related. Techniques to overcome label dependencies include grouping similar labels together [1] or modifying the loss function [3]. While they address the original issues of label similarities but remain highly specialized solution with complex training and inferencing techniques [1].

*2.2:* Text Classification under Noisy labels have been trending research topic recently and some have explored semantic similarity-based matching as an alternative to classification methods. Some bodies of work use semantic similarity between input sequences and target sequences instead of a supervised set up with noisy labels [4] whereas some other bodies of work focus on capturing a label semantic relation through Label Smoothing [5] and Label Distributions Learning [6]. There is also a direction of work to improve the loss function in a Neural Network based MLC through integrating label distributional information [7]. Yet, the work, directly comparable to ours is found in a framework named **class2simi** [8]. The authors theoretically prove that transforming *data points* with noisy labels into *data pairs* with noisy similarity labels guarantees noise reduction. They go onto build a neural network architecture that uses this transition matrix (noise data labels to noisy similarity labels) to estimate clean similarity posterior. Next, they use an inner product between clean class posterior and this approximate clean similarity posterior to learn clean class posterior. While [8] approaches the problem from an integrated Neural Network architecture point of view our approach has focused on creating an ML pipeline by reusing existing architectures and techniques. Another important difference is that class2simi is prototyped on image classification domain whereas our work is in text multi label classification.

Finally, assigning Industry Tags using Company Business Descriptions has little literature published. A reference [9] uses TF/IDF based feature engineering and SVM, NAÏVE BAYES and Fuzzy logic to make a multi label text classification does not address the problem of label relations and noises.

Hence, in a way, ours is the *first* published ML pipeline in multi label classification with noisy labels domain that uses deep learning based semantic matching with label similarity matrix and achieves robust results against gold standard ground truth.

## 3   Data

We use licensed data, thus, the examples mentioned have masked entities. Following sections discuss the features used in detail and the data preparation strategy. Our data contained Company information, their corresponding industry tags attached by a noisy labeling system and tag definitions. The logic for original tagging was unknown and during review by our SMEs they were often found to be either *partial* or *incorrect*. The sample size was about 500k.

### 3.1   Features

3.1.1 **Company Business Description (CBD)**: The textual description that describes a company's business e.g., "ABC Pvt. Ltd. is a leading payment application based out of Bengaluru. It offers financial services through its mobile application and website for customers and vendors."

3.1.2 **Industry Tag(s) (TAG):** A set of names corresponding to industry descriptions. A company can have one or more industry tags based on its businesses e.g., ABC Pvt. Ltd. from section 3.1.1 will have the industry tag as "Financial Technologies and Payment."

3.1.3 **Industry Tag Description (ITD)**: The textual description that defines a TAG e.g., description of the Tag Financial Technology can be "Financial technology companies integrate predictive behavioural analytics, data driven marketing, blockchain to provide banking services."

### 3.2   Data Preparation

We extracted a stratified (by Industry) sub sample of 2546 companies for benchmarking (benchmark set – held out). As in table 1, the companies in the data had two sets of labels, one had noisy and the other one had gold standard. It was observed that out of 2546, 668 had both the sets agreeing completely, 604 had no matches and 1274 had partial matches. Inducting from these observations, we made an expectation of *existence* of such noise distribution in the full sample. Our current work is based on this postulation. Further we prepared a *label similarity matrix* (LSM) containing N * N (N: number of labels) in which each cell can contain a 0 to 5 rating of similarity between a pair of TAGs.

## 4   Company to Industry Assignment

Our Company to Industry matching algorithm bases on two fundamental postulations –

I. Transforming data points into data pairs reduces noise [8]
II. Label correlation can be incorporated by an LSM driven data preparation for training and a corresponding choice of loss function for a semantic similarity model. This is a **novel** approach in our pipeline

Thus, we implement following steps in the pipeline (figure 1)
1. Collect a 0-5 similarity score for ~ 10% to 15% of label (TAG) pairs in LSM
2. Create triplets of CBD, ITD and Normalized Similarity Score
3. Train Sentence Transformer with a pre-trained Language Model and using Regression objective. Then infer over all ITDs and calculate performance metric
4. If model improves over past metric value, then generate embedding similarities across TAG pairs and transform the similarity values back into 0-5 rating (LSM). Next,
    o Send this LSM to SME for rating updates

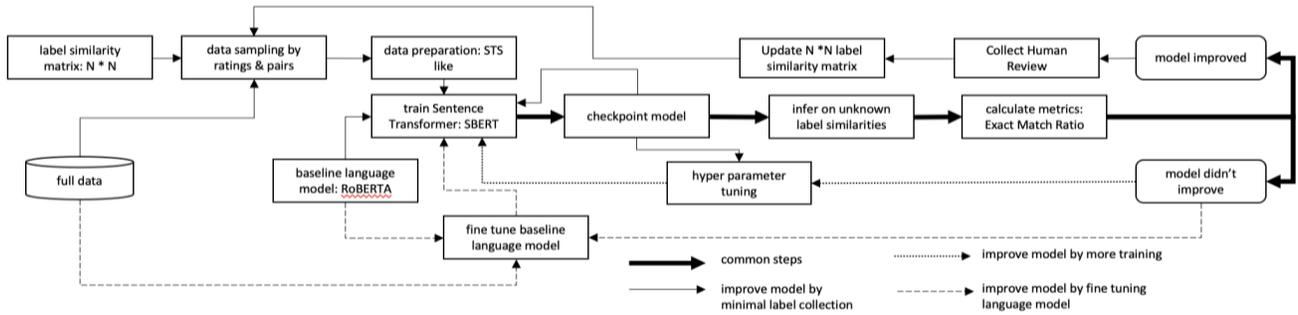

Figure 1: Flowchart of proposed system

- o Augment Training data by SME updated LSM
- o With this new data continue training
5. Else if model can still be improved by hyper parameter tuning then continue training
6. Else If model can't be improved by 4 to 6 then fine tune the language model over our data and then repeat 1 to 6

### 4.1 Minimum labeling strategy

The SMEs were given N * N LSM, and were asked to provide a score between 0 and 5 (0: no similarity, 5: complete similarity) to indicate the pairwise similarity for TAGs. At any iteration the SMEs were asked to label about a percentage of pairs through a stratification of unique combination of TAGs and rating. In this **novel** strategy SMEs have to label only ((N * (N+1))/2 – N) cases (*N much lesser than number of actual text records in a training sample*). In reality the labeling requirement is even lesser due to model generalization effects. To our observation ~15% of pair labeling is effective enough. An illustration can be found in table 2 and table 3.

|    | T1  | T2  | T3         | T4         |
|----|-----|-----|------------|------------|
| T1 | NNR | (0) | (1)        | Didn't rate |
| T2 | NNR | NNR | Didn't rate | Didn't rate |
| T3 | NNR | NNR | NNR        | Didn't rate |
| T4 | NNR | NNR | NNR        | NNR        |

Table 2: LSM before the iteration #1; NNR: not needed to rate

|    | T1  | T2     | T3      | T4  |
|----|-----|--------|---------|-----|
| T1 | NNR | [0](0) | {3} (1) | [4] |
| T2 | NNR | NNR    | {4}(3)  | [2] |
| T3 | NNR | NNR    | NNR     | [1] |
| T4 | NNR | NNR    | NNR     | NNR |

Table 3: LSM after review from SME post inference of iteration #1. [] implies model inferred and SME agreed. {} implies model inferred and SME disagreed, T: TAG

### 4.2 Semantics Similarity Matching

*One of the **novelties** lies in the postulation that a multi-label classification over dependent labels can be re-designed as a ranked labels based on their semantic similarity with input as well as among themselves*. The semantic similarity task was designed based on Sentence BERT (SBERT) [10]. SBERT intakes a record as a triplet of two texts and their similarity score in a format prescribed in STS [12]. Then it trains a Siamese network producing two discrete embeddings which then are subjected to a similarity function. The output similarity value is then regressed against ground truth scores. The loss is then used to calculate gradients which flows back along the layers and units through back propagation. It continues until the overall loss stabilizes. The network architecture is provided in figure 2. In our work, we postulate that if a pair of TAGs share a rating, then a CBD belonging to one of TAGs will share an identical rating with the ITD of the other TAG. That means we transform the LSM matrix into a description similarity matrix as shown in equation 1. The primary **novelty** we drive from this set up that a single model will see both domains of texts (i.e., CBD and ITD), thus to be capable of producing robust embeddings for both. Hence, to prepare data, we create a stratified sampling strategy to pick X (X is a finite number) different CBDs for a select ITD and a rating value pair and form triplets of (CBDs, ITD, rating). Ratings are generally normalized between 0 and 1. Therefore our SBERT model input looked like triplets of {$CBD_i$, $ITD_i$, $R_i$}. In our training, instead of BERT as shown in figure 2, we used RoBERTA [11]. Also, for specifics, we used *mean* pooling in the pooling layer.

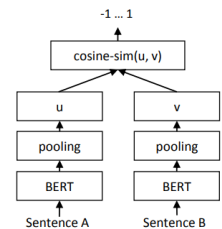

$$sim(TAG_i, TAG_j) = R_i,$$
$$CBD_i \rightarrow TAG_i,$$
$$TAG_i \rightarrow ITD_i,$$
$$CBD_j \rightarrow TAG_j,$$
$$TAG_j \rightarrow ITD_j$$
$$\Rightarrow sim(CBD_i, ITD_j) \cong R_i$$
$$R: \text{Similarity Score}$$

Figure 2　　　　　　　Equation 1

### 4.3 Regression Objective with Rated Label Similarities

It is humanely impossible to assign a precise similarity score between a pair of TAGs. So, we take 0-5 rating for Label pairs and normalize them between 0-1 and treat them as ground truth. On the other hand, the two embeddings generated for the pair of texts (CBD and ITD) are subjected to a *cosine similarity* function. Then we apply *mean square error loss* function to *regress* over the model calculated similarity value and the ground truth. This is a **novel** technique to integrate Label dependencies into the model.

### 4.4 Human Feedback & Model retraining

If model performance improves after a training iteration, we ask SMEs to update the LSM for cases where (1) model didn't agree with original entries and (2) some random cases where SME didn't label earlier. Based on this feedback we augment our training data using the stratification principle discussed in 4.2.

## 5 Results

For evaluation we compare metrics calculated using Model outcome against that of initial noisy labels, the ground truth by SMEs being common denominator.

Modeling to learn the LSM is vital as it provides scalability against newly introduced TAGs. We use Exact Match Ratio (EMR) [13] for measuring model's capability to learn the LSM. We define EMR as percentage of cases where the rating achieved by semantic similarity matches exactly with ground truth provided in the LSM. Table 4 exhibits the model's performance over 4 iterations of the pipeline and at that point model assigned correct rating for 72% of cells.

| Round # | EMR |
| --- | --- |
| 1 | 0.68 |
| 2 | 0.60 |
| 3 | 0.64 |
| 4 | 0.72 |

Table 4: EMR over LSM by SBERT model through iterations

One of the central novelties of our system is that the same model can 1) calculate label similarities among ITDs as well as to 2) infer TAGs for CBDs through similarity between ITDs and CBDs. We use average precision (AP) and average recall (AR) as performance metric in task 2. Resulting metric values are in table 5. We observe a 53% gain in recall implying model is highly performant in matching SME selected ground truths. It is to be noted that in case of precision, model always assigns 5 TAGs while noisy labeler assigns variably with median count being 1. Therefore, to gauge the predictive quality of the model we also observe average precision@K, and the result is in table 6. This shows, not only our model is 8% more precise but it does so at rank 1. This is a clear validation of our postulation that such modeling technique can be a robust replacement for a multi-label classification where labels are dependent and noisy. Another quite an important observation comes from result shown in table 7. Apparently, model is highly performant in the cases where noisy labels were totally wrong.

| Metric | Value by Model | Value by Noisy labels |
| --- | --- | --- |
| AP | 0.3027 | 0.6247 |
| AR | 0.8981 | 0.5870 |

Table 5: Aggregate metrics over benchmark set

| K | Average Precision@K | Average Similarity@K |
| --- | --- | --- |
| 1 | 0.6724 | 0.8735 |
| 2 | 0.5092 | 0.8116 |
| 3 | 0.4200 | 0.7676 |
| 4 | 0.3531 | 0.7334 |
| 5 | 0.3028 | 0.7034 |

Table 6: Average Precision@K for model over benchmark set

| Metric | Value by Model | Value by Noisy Labels |
| --- | --- | --- |
| AR | 0.9969 | 0 |
| AP | 0.2832 | 0 |

Table 7: Model metrics when noisy labels were completely incorrect

## 6 Conclusion and Future Work

In this paper we propose a novel minimal-human-in-loop pipeline that performs highly and robustly in MLC for text with noisy and dependent labels. We describe the implementation details along with comparative results. Our results exhibit an excellent generalization capability (of the model) in overall or even in most challenging highly noisy cases. Our results also exhibit a great predictive capability by the model thereby establishing the postulation that semantic similarity can be a viable alternative to MLC for texts with noisy and dependent labels. In future there is a vast scope to stress test the pipeline against various degrees of noise as well as making the components integrated into one unified model learning framework.

### Disclaimer

### ACKNOWLEDGMENTS

We would like to acknowledge the enthusiastic and cordial support received from our colleagues Steven Lau, Madison King, Nayeemur Rahman and Daniel Wu who helped us to complete our work on time and with quality.